\documentclass[conference]{IEEEtran}
\IEEEoverridecommandlockouts
\usepackage{amsmath,amssymb,amsfonts}
\usepackage{algorithmic}
\usepackage{graphicx}
\usepackage{textcomp}
\usepackage{xcolor}

\usepackage{booktabs} 
\usepackage{multirow}
\usepackage{amsfonts}
\usepackage{amssymb}
\usepackage{mathrsfs}
\usepackage{subfigure}
\usepackage{enumitem}
\usepackage{diagbox}
\newtheorem{definition}{Definition}
\usepackage{filecontents}
\usepackage[noadjust]{cite}

\usepackage[ruled,linesnumbered]{algorithm2e}

\def\BibTeX{{\rm B\kern-.05em{\sc i\kern-.025em b}\kern-.08em
    T\kern-.1667em\lower.7ex\hbox{E}\kern-.125emX}}
\begin{document}

\title{GCN-ALP: Addressing Matching Collisions in Anchor Link Prediction}

\author{\IEEEauthorblockN{Hao Gao, Yongqing Wang, Shanshan Lyu,  Huawei Shen, Xueqi Cheng}
\IEEEauthorblockA{\textit{CAS Key Laboratory of Network Data Science and Technology}\\
\textit{Institute of Computing Technology, Chinese Academy of Sciences}\\
Beijing, China \\
\{gaohao, wangyongqing, lvshanshan, shenhuawei, cxq\}@ict.ac.cn}
}

\maketitle

\begin{abstract}
Nowadays online users prefer to join multiple social media for the purpose of socialized online service. The problem \textit{anchor link prediction} is formalized to link user data with the common ground on user profile, content and network structure across social networks. Most of the traditional works concentrated on learning matching function with explicit or implicit features on observed user data. However, the low quality of observed user data confuses the judgment on anchor links, resulting in the matching collision problem in practice. In this paper, we explore local structure consistency and then construct a matching graph in order to circumvent matching collisions. Furthermore, we propose graph convolution networks with mini-batch strategy, efficiently solving anchor link prediction on matching graph. The experimental results on three real application scenarios show the great potentials of our proposed method in both prediction accuracy and efficiency. In addition, the visualization of learned embeddings provides us a qualitative way to understand the inference of anchor links on the matching graph.
\end{abstract}

\section{Introduction}\label{introduction}
With the benefit of socialized online service, people are used to taking activities across multiple social media simultaneously. For example, Facebook users would share their travel journals with attached photographs shared in Flickr. According to the research report\footnote{http://www.pewinternet.org/2018/03/01/social-media-use-in-2018/}, the median of using major social platforms of American are three in 2018. Therefore, the linkage of user data (also be referred with \textit{anchor link}) across social networks becomes a key issue on better understanding users to promote various of applications, e.g., recommendation ~\cite{Li2014, man2017cross} and user profiling~\cite{Li2014a}. 

The problem \textit{anchor link prediction} (ALP) is formalized to link user data with the common ground on user profile, content and network structure across social networks. Most of the traditional works~\cite{Liu2013,Zafarani2009,Zafarani2013, Nie2016Identifying} concentrated on learning matching function with explicit or implicit features on observed user data. For example, MOBIUS introduced a mapping function by exploiting users' unique behavior patterns in user names~\cite{Zafarani2013}. Riederer et al. developed an efficient approach to quantize the uniqueness of user activities in users' trajectory data~\cite{Riederer2016Linking}. Man et al. proposed a two-phase mapping method. The mapping function is learned by a multi-layer perceptron as the inputs of implicit structural features resulted from network embedding~\cite{Man2016}. Moreover, some literatures constructed mapping function with integrated user data~\cite{Zhang2018, Zhong2018}. The consensus in these literatures is to capture consistent and unique representations of users across networks, improving the prediction performance in ALP. However, the low quality of observed user data confuses the judgment on anchor links, resulting in many plausible matched identities in practice (i.e., matching collisions). 

\begin{figure}[h]
	\tiny
	\centering
	\includegraphics[scale=0.25]{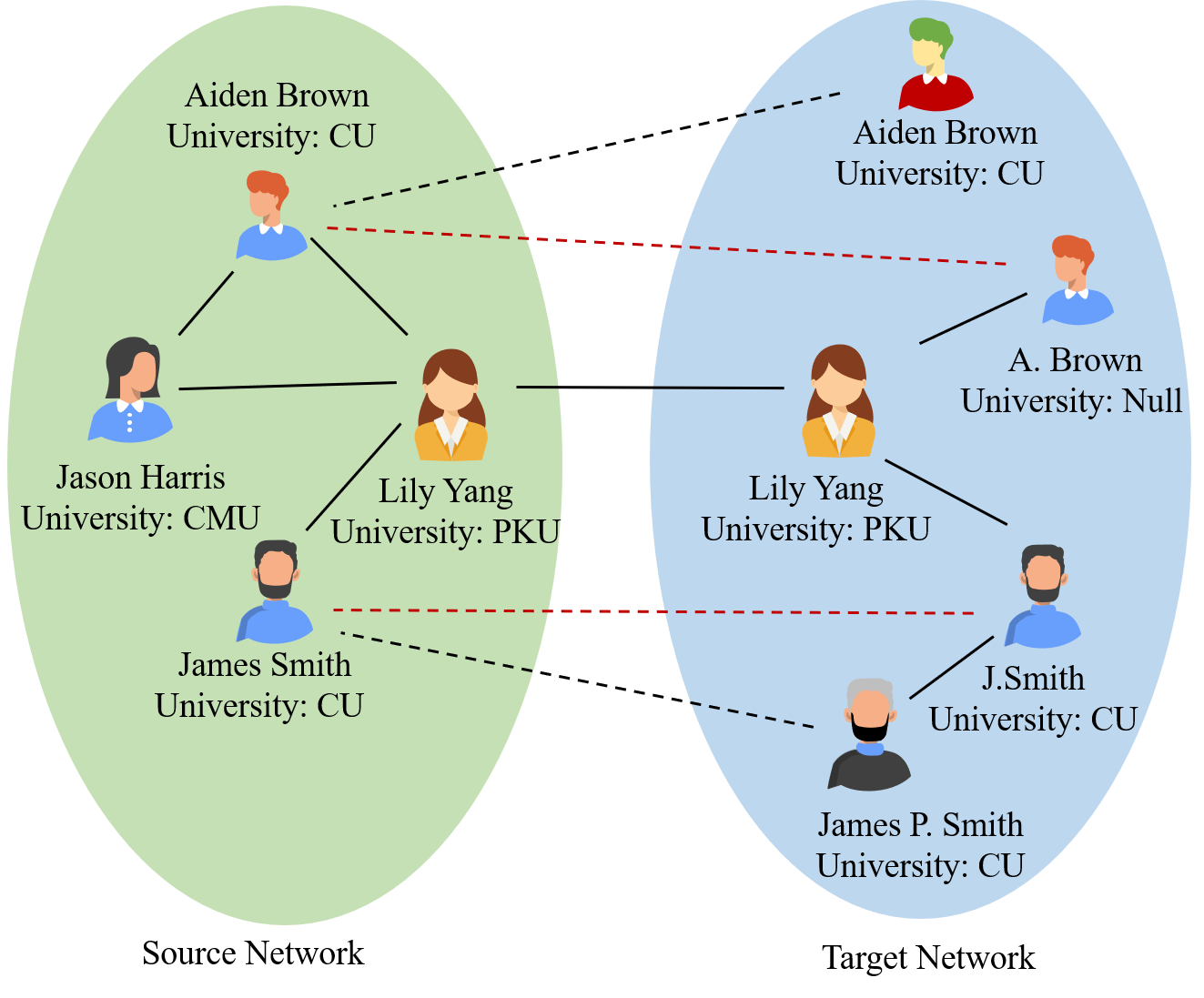}
	\caption{Illustration of matching collisions in ALP. The solid lines are the relationships between users in the source or target networks. The dotted lines refer to the inferred candidates for ALP. The users connected with the \textbf{red} dotted lines indicate that they are the same users (i.e. anchor links).} \label{demo}
\end{figure} 
The matching collision problem implies that users from different networks have the same or similar markers regardless of anchors or not. Indeed, such kinds of matching collisions are general in real applications. In particular, we can take examples on exploring anchor links over name and college information of users. As shown in Figure~\ref{demo}, ``James Smith” is a college student at Cornell University, registered his account in the source network. Meanwhile, ``J. Smith” and ``James P. Smith” seem like the best matchups for similar properties (similar name and university) in the target network. In this case, the clues from name and college similarity may incur confused matchups in determining the true correspondence. Moreover, the overreliance on single user property may also lead wrong matchups in ALP. For instance, ``Aiden Brown” in the source network has two candidate correspondences in the target network, i.e. ``Aiden Brown” and ``A. Brown”. It seems like that “Aiden Brown” in target network should be the best matchup when considering user name. However, the “A. Brown” is the true correspondence for her. \emph{It is convenient for users who can change their properties in social networks, leading a great number of matching collision problems in practice.} To further examine the problem, we take the statistics on three real application scenarios, including Douban online-offline, Flickr-Lastfm, and Flickr-Myspace where the features for ALP are well-constructed by literatures~\cite{Zhang2016}. We found every observed anchor node has 1 matching collision node at least, amounting to over 11\%, 18\%, and 22\% matching collision pairs in three published datasets respectively.

It motivates us that we can introduce more reliable information to reduce the number of matching collisions, improving the prediction accuracy in ALP. Thus, we further examine the datasets and explore \emph{local structure consistency}. The literatures~\cite{Liu2016,Zhang2016} explained the local structure consistency that users would prefer to transfer their neighborhoods to another social network, that is, the candidates who have true anchors in their neighborhoods would be the more possible inference than others. In our preliminary experiments, we can find that nearly 62\% of 1-hop neighbors are preserved across Douban Online-offline networks, and the percentage is over 32\% and 12\% in the scenarios of Flickr-Lastfm and Flickr-Myspace respectively. In this way, local structure consistency gives an important hint for predicting anchor links.

To better utilize structural information across networks, in this paper, we construct a matching graph, following the local structure consistency, where the nodes and edges are unions of two networks. The proposed matching graph promise that the inference of anchor links is followed by both similarity of user characteristics and the relationships to labeled anchor links. Furthermore, we introduce graph convolution networks (GCN) for resolving ALP in the matching graph. The GCN approach provides a way to convolve local neighbors and predict the node labels by transductive inference, utilizing the local structure consistency on the matching graph. As for the great size of matching graph, we further propose a mini-batch strategy to accelerate convolution on matching graph. More specifically, the main contributions of our work are three folds:
\begin{itemize}
	\item We introduce matching collision problem in ALP, which is an important issue in real applications but ignored by most previous works.
	\item We explore local structure consistency for reducing matching collisions. Furthermore, we construct a matching graph, following the local structure consistency. The proposed matching graph promises that the inference of anchor links is subjected to both similarities of user characteristics and the relationships to labeled anchor links.
	\item We propose graph convolution networks to solve ALP in matching graph, where the mini-batch strategy is proposed to circumvent the scalability of the combination of the networks. The experimental results on three real application scenarios show the great potentials of our proposed method in both prediction accuracy and efficiency. In addition, the visualization of learned embedding provides us a qualitative way to understand the inference of anchor links on matching graph.
\end{itemize}	
\section{Related Work}\label{Related Work}
\subsection{Anchor Link Prediction}

Based on the information leveraged to predict anchor links, existing methods are categorized to profile-based methods, network-based methods and hybrid methods. 

Intuitively, users in different social networks tend to have similar profiles~\cite{Liu2013,Zafarani2009,Zafarani2013}. For example, Zafarani et al. model the users' behaviors on different social networks based on the user names, e.g. edit distance, typing patterns, etc~\cite{Zafarani2013}. Besides, researchers exploit other attributes to predict anchor links such as the trajectory of locations~\cite{Riederer2016Linking}, the content of users' posts~\cite{Nie2016Identifying}. However, users' attributes could be missing or suffer from collisions in different social networks in practice. 

Network-based methods are categorized into two main manners: unsupervised way and supervised way. The objective of unsupervised methods is to preserve the overlap of the source and target network as much as possible. Klau et al. propose an optimization formulation to maximize the structural matching score to align two networks~\cite{Klau2009}. Koutra et al. propose to learn the sparse and real-valued permutation matrix to align networks~\cite{Koutra2013BIG}. However, the unsupervised methods fail to leverage observed anchor nodes' information. An intuitive supervised way is to extract structural features, and features are fed to a classifier to predict whether a pair of nodes is an anchor link or not~\cite{Kong2013}. With the popularity of deep learning, researchers study how to propagate the proximity of anchor links through networks via an embedding learning process. Man et al. propose to learn the embeddings of both networks independently and then map the source embedding to the target embedding~\cite{Man2016}. Liu et al. unify these two stages to learn better embeddings~\cite{Liu2016}. To accelerate the matching process, Wang et al. propose to learn the binary representations of users~\cite{wang2019learning}. Then the  hashing-based search algorithms are applied to infer anchor links. Network-based methods neglect the users' attributes, which may be difficult to identify users when the network topology is different across social media.	

Recently some hybrid methods are proposed. Kong et al. propose to extract social features with supervised information to train a classifier and matched anchor nodes by solving "stable marriage problem"~\cite{Kong2013}. However, it neglects the intrinsic network structures and is sensitive to noises of features. Liu et al. propose a multi-objective optimization framework composed of both attributes and cross-network structure consistency~\cite{Liu2014HYDRA}. Zhang et al. propose COSNET~\cite{Zhang2015}, in which energy-based functions are modeled incorporating local consistency and global consistency. Zhong et al. propose an unsupervised framework that co-train both attributes and network features iteratively~\cite{Zhong2018}. However, the proposed methods fail to take the matching collisions into consideration.
\subsection{Graph Neural Networks}

An interesting work in graph neural network is rising which integrates local node features and network topology in the convolutional ways~\cite{Lee2018, Velickovic2018,Xu2019, xu2019graph}. The key idea behind GCNs is that the convolution operation smooths the neighbors of the center node, which makes the nodes with similar features in the same cluster. Thus it is easier to classify the unlabeled samples. Niepert et al. analogize the traditional convolution to graph data, where locally connected regions are convolved~\cite{Niepert2016Learning}. However, It's heuristic and time-consuming to determine the neighbors of nodes. Bruna et al. propose to convolve networks based on graph laplacian in the spectrum domain and learn a diagonal matrix of parameters which have the same size of the number of nodes in networks ~\cite{Bruna2014Spectral}. Defferrard et al. reduce the parameters learned from the previous work, where the parameters in the diagonal matrix are parameterized~\cite{Defferrard2016Convolutional}. Kipf et al. further simplify the parameters in ~\cite{Kipf2016}. In their pilot work, they apply GCNs to semi-supervised node classification. However, these methods are not scalable for large graphs. Hamilton et al. propose GraphSAGE~\cite{hamilton2017inductive} that learns the aggregating function to generate nodes' embeddings from their local neighbors incorporating the nodes' attributes.
\section{Problem Formulation}\label{Problem Formulation}
Let $\mathcal{G}^s$ and $\mathcal{G}^t$ be the two unweighted and undirected graphs. The source network is denoted as $\mathcal{G}^s=(\mathcal{V}^s, \mathcal{E}^s, \mathbf{X}^s)$. The set $\mathcal{V}^s$ is the set of nodes in source network, containing $n^s$ nodes. The set $\mathcal{E}^s\subset \mathcal{V}^s \times \mathcal{V}^s$ is the set of edges among nodes in $G^s$. The matrix $\mathbf{X}^s=[\mathbf{x}^s_1, \mathbf{x}^s_2, ..., \mathbf{x}^s_n]$ is the matrix of nodes' attributes, and a column $\mathbf{x}^s_j$ is a feature vector of node $j$. Similarly, the target network is denoted as $\mathcal{G}^t=(\mathcal{V}^t, \mathcal{E}^t, \mathbf{X}^t)$ which contains $n^t$ node. Traditionally, the constructed mapping function aims to learn the consistency between two nodes across networks according to $\mathbf{X}^s$ and $\mathbf{X}^t$. However, these methods suffer the matching collision problem under indiscriminative user features. To circumvent the problem, we firstly introduce a matching graph jointly constructed by source and target networks. The objective of the matching graph is to promise the \textit{local structure consistency} across networks. Here we introduce the definitions on the matching graph.

\begin{definition}\label{matching_node}
	\textbf{Matching Nodes.} We define the matching node $v_i u_j$, constituted by node $v_i \in \mathcal{E}^s$ in source network $\mathcal{G}^s$ and node $u_j \in \mathcal{E}^t$ in target network $\mathcal{G}^t$. The attributes of matching node are calculated by the joint attributes on $v_i$ and $u_j$. We denote the matching nodes as $\mathcal{V}^m=\{v_i u_j| \forall v_i \in \mathcal{V}^s,\forall u_j \in \mathcal{V}^t\}$, and the attributes matrix of matching nodes is $\mathbf{X}^m=[\mathbf{x}^m_1, \mathbf{x}^m_2, ..., \mathbf{x}^m_{n^s*n^t}]$, where $\mathbf{x}^m_{i*j} = \theta(\mathbf{x}_i^s, \mathbf{x}_j^t) $ and $\theta$ is a task-specific attribute calculator function as inputs of $\mathbf{x}_i^s$ and $\mathbf{x}_j^t$.
\end{definition} 
Generally, traditional ALP problem merely define the function $\theta$ to measure the similarity between nodes in source and target networks. However, the massive matching collisions mislead the measurements when the inputs are less discriminative on ALP task. With consideration of local structure consistency, we introduce matching edges between matching nodes, reflecting common neighbors across networks. The definition of matching edge is depicted as follows.
\begin{definition}\label{matching_edge}
	\textbf{Matching Edges.} Given two matching nodes $v_i u_j$ and $v_k u_l$, the matching edge holds if $(v_i, v_k)$ and $(u_j, u_l)$ are both directly linked in network $\mathcal{G}^s$ and $\mathcal{G}^t$ respectively. Formally, the matching edges are the set $\mathcal{E}^m=\{(v_i u_j, v_k u_l)|\forall(v_i, v_k) \in \mathcal{E}^s, \forall (u_j, u_l) \in \mathcal{E}^t \}$.
\end{definition}

\begin{figure*}
	\includegraphics[scale=0.4]{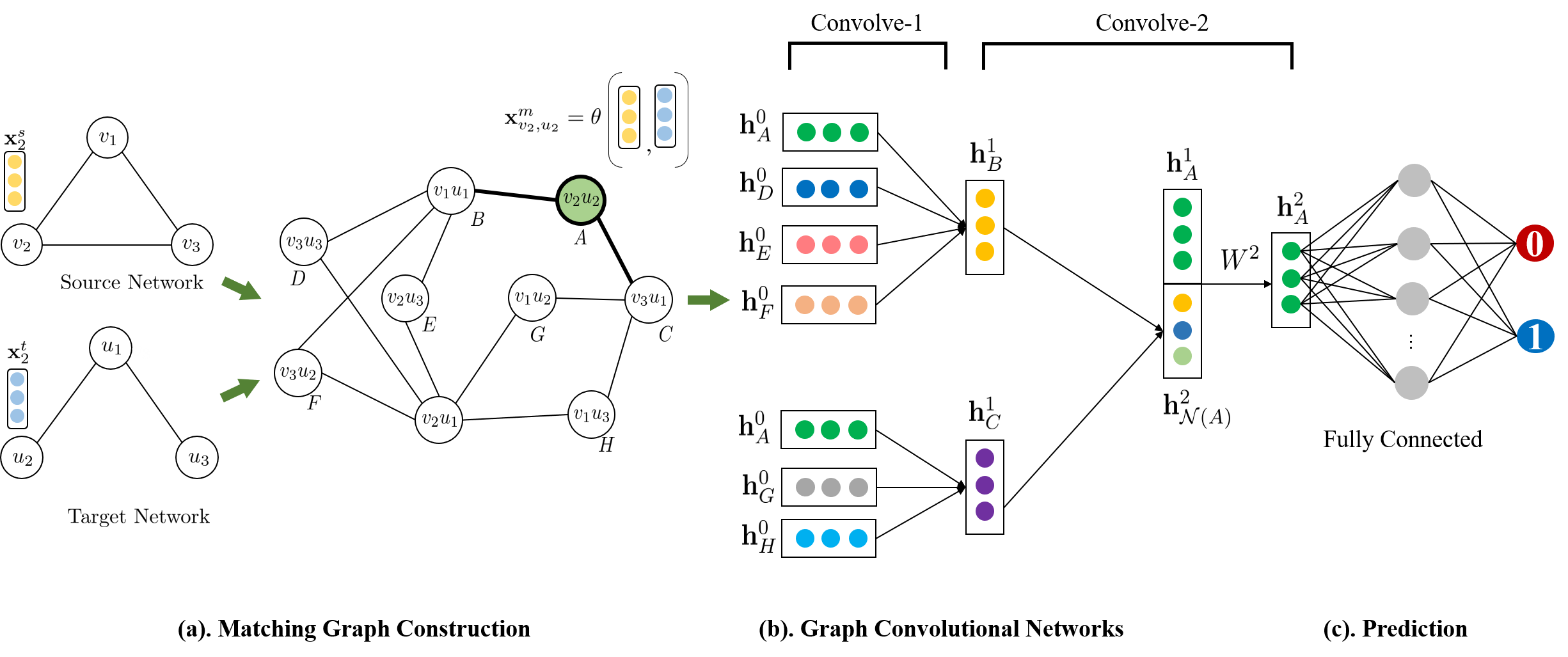}
	\centering
	\caption{Framework of our proposed model. (a). Matching Graph Construction. The matching graph is constructed based on definition~\ref{matching graph}. (b). Graph convolution networks learn the representations of matching node A. (c). Prediction. The representations $\mathbf{h}^{2}_{A}$ are fed to a fully connected layer to predict whether A is an anchor link.}
	\label{framework}
	\vskip -10pt
\end{figure*}

With matching edges, the similarity defined on matching nodes can be passed, providing an inference regularization on ALP. Based on the defined matching nodes and matching edges above, the matching graph can be formalized as follows. 
\begin{definition}\label{matching graph}
	\textbf{Matching graph.} The matching graph is a graph with \textit{matching nodes} and \textit{matching edges}. We denote \textit{matching graph} as $\mathcal{G}^m=(\mathcal{V}^m, \mathcal{E}^m, \mathbf{X}^m)$.
\end{definition}
%
The construction of matching graph is illustrated in Fig.~\ref{framework}(a). According to the definition~\ref{matching_node}, we combine the nodes in source and target networks, generating $3\times 3$ matching nodes. Then we set linkages between match nodes. More specifically, we can discuss the linkage around matching node $v_2 u_2$ as an example. The matching node is composed of node $v_2$ and $u_2$ in source and target networks. The node $v_2$ has two direct neighbors in source network, i.e., node $v_1$ and $v_3$, while node $u_2$ has one direct neighbor in target network, i.e., node $u_1$. According to the definition~\ref{matching_edge}, the direct neighbors of matching node $v_2 u_2$ can be specified to $v_1 u_1$ and $v_3 u_1$.  
\section{Methodology}\label{Method}
The proposed matching graph is the basis of the solution on matching collision problem, supporting the inference of anchor links which can be followed by both similarity of user characteristics and the closeness to labeled anchor links. Meanwhile, the GCN approach provides a way to convolve local neighbors and predict the node labels by transductive inference. Therefore, we propose GCN on the matching graph to leverage the benefits on both matching graph and graph convolution. Furthermore, we propose a mini-batch strategy to overcome the huge computational cost caused by the size of the matching graph.

In the following, we will introduce the model architecture at first and then give our efficient learning strategy.

\subsection{Model Architecture}\label{original_model_section}

According to the idea of graph convolution, we can generate node representations by the layer-wise convolving operation. The convolving operation and the proposed model architecture are depicted in Figure~\ref{framework}. 

\subsubsection{Representation generation}
We firstly describe a forward process for node representation learning. The learning process is described in Figure~\ref{framework}(b). To better illustrate the learning process, we use the matching node A as a specific case. Firstly, we set $K$ convolution layers, aggregating node attributes around the center node. In the first convolution layer, the convolving function aggregates representations on 1-hop neighbors of node $A$ (Abbreviated for $v_2u_2$), i.e., node $B$ ($v_1u_1$) and $C$ ($v_3u_1$). The node representations of B and C are iteratively learned by the second convolution layer, aggregating representations on 2-hop neighbors of node $A$. For the convenience of implementation, we set the $(K-i+1)$-th layer to procedure the convolving operation on $i$-hop neighbors of center node. In general, the representation learning of node $v$ on $k$-th layer, $\mathbf{h}_v^k$, can be formalized by a non-linear convolving transformation, that is,
$$\mathbf{h}^k_v \leftarrow \sigma(\mathbf{W}^k\cdot CONCAT(\mathbf{h}^{k-1}_{v}, \mathbf{h}^k_{\mathcal{N}(v)}))$$
where $\mathbf{W}^k$ means the weight matrices in layer $k$, $\sigma$ refers to a sigmoid function and \textit{CONCAT} refers to concatenation of $\mathbf{h}_v^{k-1}$ and $\mathbf{h}^k_{\mathcal{N}(v)}$. The neighbor representation $\mathbf{h}^k_{\mathcal{N}(v)}$ is calculated by aggregator function as follows,
%
$$
\mathbf{h}^k_{\mathcal{N}(v)} \leftarrow \textit{AGG}_k ( \{\mathbf{h}^{k-1}_u, u\in \mathcal{N}(v) \})
$$
In our Algorithm, we set $AGG$ as mean function, which aggregates the average representations of nodes as the hidden representations.
\subsubsection{Objective function}
After the node representation learning, we apply a fully connected layer to predict whether the matching node is a pair of anchor link. We define the loss function by cross-entropy on the fully connective layer:
$$L=\frac{1}{n}\sum_{v=1}^{n}CrossEntropy(f(\mathbf{z}_v), y_v)$$	

\begin{algorithm}[htb]
	\SetKwInOut{Input}{Input}
	\SetKwInOut{Output}{Output}
	\SetKwFunction{Batch-sampling}{Batch-sampling}
	\caption{Mini-batch Embedding Generation.\label{alg_embedding_minibatch} } 
	\Input{Source network $\mathcal{G}^s=(\mathcal{V}^s, \mathcal{E}^s, \mathbf{X}^s)$;
		Target network $\mathcal{G}^t=(\mathcal{V}^t, \mathcal{E}^t, \mathbf{X}^t)$;
		Set of nodes $\mathcal{M} \in \mathcal{V}^m$ on matching graph;
		Number of convolution layers $K$;
		Number of sampling nodes $q$;
		Weight matrix $\mathbf{W}_k, k\in\{1,...K\}$;
		aggregator functions $AGG_k, k\in\{1,...K\}$}
	\Output{Vector representation $\mathbf{z}_v$ for each node $v$ in $\mathcal{M}$}
	\tcc{Batch sampling in source and target network}
	$\mathcal{B}^K \leftarrow \mathcal{M}$\;
	\For{$k=(K-1), ..., 0$}{
		$\mathcal{B}^k \leftarrow$ Batch-sampling$(\mathcal{B}^{(k+1)},\mathcal{G}^s, \mathcal{G}^t, q)$\tcp*[h]{ See Algorithm~\ref{alg_batch_sample}}\;
	}
	
	$\mathbf{h}_{v_i u_j}^0 \leftarrow \theta(\mathbf{x}_{v_i}, \mathbf{x}_{u_j}), \forall v_i u_j\in \mathcal{B}^0 $\;
	\For{$k=1, ..., K$}{			
		\For{$v_i u_j\in \mathcal{B}^k$}{
			$\mathcal{N}(v_i u_j) \leftarrow 	\{v_k u_l|\forall v_k \in \mathcal{N}(v_i) ,\forall u_l \in \mathcal{N}(u_j) \}$\;\label{8}
			$\mathbf{h}_{\mathcal{N}(v_i u_j)}^k \leftarrow AGG_k(\{h_{a'}^{k-1}, \forall a' \in \mathcal{N}(v_i u_j)\})$\;\label{9}
			$\mathbf{h}_{v_i u_j}^k \leftarrow \sigma(\mathbf{W}^k \cdot CONCAT(\mathbf{h}_{v_i u_j}^{k-1}, \mathbf{h}_{\mathcal{N}(v_i u_j)}^k))$\;\label{10}
			$\mathbf{h}_{v_i u_j}^{k} \leftarrow \frac{\mathbf{h}_{v_i u_j}^{k}}{\|\mathbf{h}_{v_i u_j}^{k}\|_2}$\;	
		}
	}\label{2}

	$\mathbf{z}_v \leftarrow h_v^{K}, \forall v \in \mathcal{M}$;
	
\end{algorithm}

\begin{algorithm}[htb]
	\SetKwInOut{Input}{Input}
	\SetKwInOut{Output}{Output}
	
	\SetKwInOut{Input}{Input}
	\SetKwInOut{Output}{Output}
	
	\caption{Batch-sampling.\label{alg_batch_sample}} 
	\Input{Batch $\mathcal{B}^k$;
		Source network $\mathcal{G}^s=(\mathcal{V}^s, \mathcal{E}^s, \mathbf{X}^s)$;
		Target network $\mathcal{G}^t=(\mathcal{V}^t, \mathcal{E}^t, \mathbf{X}^t)$;
		Number of sampling nodes $q$;
	}
	\Output{Batch of next layer $\mathcal{B}^{k-1}$}
	$\mathcal{B}^{k-1} \leftarrow \mathcal{B}^{k}$\;
	\For {$\forall uv \in \mathcal{B}^k$}{
		$Nei\_nodes=[]$\;
		
		\For{$\forall k \in \mathcal{N}(u)$}{
			\For{$\forall l \in \mathcal{N}(v)$}{
				$Nei\_nodes \leftarrow Nei\_nodes+kl$\;\tcp*[h]{Construct matching node.}
			}
		}
		$Nei\_nodes = sample(Nei\_nodes,q)$\tcp*[h]{Sample $q$ nodes from $Nei\_nodes$} \label{line_sample_strategy}\;
		$\mathcal{B}^{k-1} \leftarrow \mathcal{B}^{k-1} + Nei\_nodes$\;
	}
	\Return $\mathcal{B}^{k-1}$\;
\end{algorithm}
\subsection{Optimization}\label{mini-batch}
According to the illustration in Section~\ref{Problem Formulation}, the computational complexity is highly dependent on matching graph construction. The number of nodes and edges is up to $|V^s| * |V^t|$ and $|E^s| * |E^t|$ in a complete matching graph, respectively, both growing in a quadratic way. The huge network size leads to the intolerable storage and computation cost in practice. Therefore, the proposed mini-batch strategy aims to reduce the complexity in both storage and computation. In particular, we adopt a dynamic way to construct matching graph during batch sampling. For learning the representation of target nodes at K-th layer, we will sample its neighborhoods and aggregate their attributes at previous layer. The process can be repeated until the layer in beginning. In every iteration of batch sampling, we merely sample a limited number of target nodes, i.e., 128, 256, 512. Furthermore, we also fix the number of sampled neighborhoods by bootstrap sample. In this way, the storage and computation cost is constant in mini-batch training process, circumventing the complexity problem in practice. 
Meanwhile, the neighborhood sampling is discussed in our experiments, that is, random sampling and feature importance sampling. In random sampling, the neighborhoods are sampled by a uniform distribution. In feature importance sampling, we select the neighbors according to node similarity, evaluated by node attributes. Overall, the dynamic process of matching node construction and batch sampling are depicted in Algorithm~\ref{alg_batch_sample}, and the complete learning process is presented in Algorithm~\ref{alg_embedding_minibatch}.

\section{Experiments}\label{Experiments}
In this section, we conduct experiments to evaluate the effectiveness of our model and compare it with several state-of-the-art methods. Moreover, we evaluate the efficiency of our approach, demonstrating that our model is scalable. Finally, we visualize the embeddings learned by our model, providing us a qualitative way to understand the inference of anchor links on matching graph.

\subsection{Experiment Settings}
We introduce the settings of the experiments in this section.
\subsubsection{Datasets}
We employ four SNS networks from~\cite{Zhang2016} to demonstrate the effectiveness of our approach. The ground-truth is already collected by the datasets. The details of the datasets are described as follows:
\begin{itemize}
	\item{\textbf{D1. Douban online vs. Douban offline.}} The  offline network has 1,118 users and the online network has 3,906 users which contain all users in  the offline network. The attributes of nodes are the users' locations. The ground-truth is collected from the webpage, which contains 1,118 observed anchor links.
	\item{\textbf{D2. Flickr vs. Lastfm.}} Flickr network and Lastfm network have 12,974 and 15,436 users respectively. We use genders as users' attributes. Besides, we classify the node to three categories based on their Pagerank score: the 1\% highest nodes are ``opinion leader'', the next 10\% are "middle class" and the rest of users are "ordinary users". The ground-truth contains 425 observed anchor links.
	\item{\textbf{D3. Flickr vs. Myspace.}} Flickr and Myspace have 6,714 and 10,733 users respectively. We extract attributes of users in the same way for D2. The ground-truth contains 267 observed anchor links. 
	
\end{itemize}
\subsubsection{Comparison Methods}We compare our method with the state-of-the-art methods, which are briefly described below. 

\begin{itemize}
	\item \textbf{Attribute Classification (AC)} trains a SVM classifier on the concatenation of users' attributes.
	
	\item \textbf{MNA~\cite{Kong2013}} extracts pairwise network features and heterogeneous features to train a classifier, and infer the anchor links by solving ``table marriage problem'' to find the global matching of anchor nodes. 
	
	\item \textbf{PALE~\cite{Man2016}} is a two-stage embedding and mapping method that learns the representations of two networks separately and then learns a mapping function to predict anchor links. In order to compare the PALE with our method, we add attribute features in the second stage, i.e. concatenating the feature with the network embeddings to learn mapping function.
	
	\item \textbf{Label Propagation (LP)~\cite{Xiaojin2002}} is an iterative matrix production on the hyper-graph to propagate the labels based on the similarities between nodes. We set the weight of the edges as the cosine similarities($1-cosine\_distance$) between the attribute features of nodes.
	\item \textbf{Mego2vec~\cite{Zhang2018}} is an end-to-end method to predict anchor links. The authors propose to learn the embeddings of users based on their features and network topology. Finally, a graph convolution networks are constructed to propagate the features to predict the anchor links. 
\end{itemize}

\subsubsection{Evaluation Metrics}We adopt two evaluation metrics to compare our method with baselines, which are $MRR$(Mean Reciprocal Rank) and $Hits@R$. $MRR$ is given as:
$\label{MRR}
	MRR=\frac{1}{n}\sum_{i=1}^{n}\frac{1}{r_i}
$
, where $r_i$ is the rank of anchor node $i$ in the candidates of the target network. Besides, $Hits@R$ is given as:
$\label{ACC}
	Hits@R=\frac{1}{n}\sum_{i=1}^{n} \mathbb{I}(r_i<=R)
$
, where $\mathbb{I}$ is an indicator function, and $r_i$ is the rank of anchor node $i$ in the candidates of the target network. We set $R$ as 1 and 10 in our experiments. 
\subsubsection{Implementation Details}
We partition the labeled data into two parts for training and test. The training set is randomly sampled by 10\%, 50\% and 80\% percentage from labeled data and rest data is used for testing. Moreover, we randomly sample a set of non-anchor links as negative samples where the ratio of positive and negative samples is 1:1. To accelerate the evaluation, for each source anchor node in the test set, we choose $n$ non-anchor nodes together with the anchor node in the target network as the candidates. We set $n$ as 20, and thus 21 candidates nodes are chosen in the experiments. In our experiments, we set the function $\theta$ as the cosine similarity, batch size as 128, learning rate as 0.01, embedding size as 128. The number of hops is set as 2. Besides, the numbers of the neighbors sampled in both 1-hop and 2-hop are set as 5 and 10. We implement the models in TensorFlow with Adam optimizer on the server with K80 GPUs. All the baselines are tuned with the optimal parameters. Besides, LP is not salable when the networks are large due to constructing the whole matching graph. Thus the result of LP on Flickr-Lastfm is not shown.

\begin{table*}[]
	\caption{Prediction results of different methods while training ratio grows.}
	\label{exp_result}
	\centering
	\begin{tabular}{cc|ccc|ccc|ccc}
		\hline
		\multicolumn{2}{c|}{\textbf{Training Ratio}}                                                 & \textbf{}      & \textbf{0.1}   & \textbf{}           & \textbf{}      & \textbf{0.5}   & \textbf{}      
		& \textbf{}      & \textbf{0.8}   & \textbf{}      \\ \hline
		\textbf{Dataset}                                                           & \textbf{Method} & MRR            & Hits@1         & Hits@10        & MRR            & Hits@1         & Hits@10        & MRR             & Hits@1         & Hits@10        \\ \hline
		\multirow{5}{*}{\begin{tabular}[c]{@{}c@{}}Online-\\ Offline\end{tabular}} & MNA             & 0.050          & 0.000          & 0.000          &  0.808          & 0.701          & 1.000 & 0.856 &
		0.775 &	1.0 \\
		& PALE           & 0.185          & 0.049          & 0.520          & 0.174          & 0.056          & 0.446          & 0.193          & 0.069		   & 0.473\\
		& AC               & 0.695          & 0.531          & 1.000 & 0.693          & 0.522          & 1.000 &  0.709 &0.554 & 1.0 \\
		& LP              & 0.181          & 0.060          & 0.446           & 0.221          & 0.109          & 0.424          &0.247 &0.127 & 0.457\\
		& Mego2vec              & 0.608          & 0.400          & 1.000          & 0.600          & 0.370          & 0.996          & 0.600 & 0.393&1.0\\
		& Ours            & \textbf{0.804} & \textbf{0.681} & \textbf{1.000} & \textbf{0.811} & \textbf{0.683} & \textbf{1.000}& \textbf{0.813} & \textbf{0.692} & \textbf{1.000}\\
		\hline
		\multirow{5}{*}{\begin{tabular}[c]{@{}c@{}}Flickr-\\ Lastfm\end{tabular}}  & MNA             & 0.172          & 0.050          & 0.475          & 0.171          & 0.055          & 0.481          &0.151
		&0.017 & 0.475\\
		& PALE           & 0.158          & 0.028          & 0.470         & 0.188          & 0.050          & 0.486          & 0.155 & 0.028 & 0.453
		\\
		& AC              & 0.238          & 0.105          & 0.547         & 0.224          & 0.077          & 0.575          &0.238 & 0.099 & 0.569\\
		& LP             & -              & -              & -              & -              & -              & -              & -              & -              & -              \\
		& Mego2vec              & 0.232          & 0.099         & 0.552         & 0.228          & 0.077         & 0.597          & 0.240          & 0.105         & 0.558          \\
		& Ours            & \textbf{0.254} & \textbf{0.116} & \textbf{0.569} & \textbf{0.266} & \textbf{0.116} & \textbf{0.624}& \textbf{0.269} & \textbf{0.120} & \textbf{0.631} \\ \hline
		\multirow{5}{*}{\begin{tabular}[c]{@{}c@{}}Flickr-\\ Myspace\end{tabular}} & MNA             & 0.129          & 0.028          & 0.402          & 0.122          & 0.019          & 0.393       & 0.120 &
		0.0 &		0.449
		\\
		& PALE            & 0.190          & 0.047          & 0.505         & 0.165          & 0.028          & 0.458          & 0.181 & 0.037 & \textbf{0.542}
		\\
		& AC              & 0.135          & 0.019          & 0.402          & 0.182          & 0.037          & 0.467           & 0.184          & 0.037          & 0.533          \\
		& LP              & 0.176          & 0.037          & 0.486         & 0.157          & 0.028          & 0.458         & 0.163          & 0.028          & 0.467 \\
		& Mego2vec              & 0.198          & 0.047          & 0.495         & 0.198          & 0.075          & 0.493         & 0.202          & 0.079          & 0.503          \\
		& Ours           & \textbf{0.214} & \textbf{0.075} & \textbf{0.505} & \textbf{0.228} & \textbf{0.084} & \textbf{0.505}  & \textbf{0.230} & \textbf{0.085} & 0.509 \\ \hline
	\end{tabular}
\end{table*}
\subsection{Performance Comparison}
In this section, we compare the performance of our model with the chosen state-of-the-art methods to validate the effectiveness of our method. Table~\ref{exp_result} presents the overall experimental results. 
\begin{itemize}
	\item The overall performance of Flickr-Lastfm and Flickr-Myspace are worse than the score of the Online-Offline dataset which is because more matching collisions exist in the former dataset as we analyze in Section~\ref{introduction}.
	\item  Our proposed approach performs better than the baseline method AC which merely utilizes the user attribute information, indicating that our model is effective in predicting anchor links by constructing a matching graph with joint user attributes and local neighborhoods.
	\item  Our method outperforms MNA which extracts the attributes and network statistics features to predict anchor links. It proves that predicting the node labels by transductive inference from the local structure consistency on matching graph is more effective than the indiscriminate network statistics features by MNA. We notice that MNA is not working when the training ratio is low in the Online-Offline dataset due to the indiscriminate network statistics features based on the rarely observed anchors. Besides, our method is also better than PALE and LP which fail to model the local structure consistency.
	\item  Our methods are scalable compared to method LP. LP needs to construct the matching graph and then propagate labels on it, such that it is not scalable on big datasets like Flickr-Myspace. 
	\item  Our approach performs better than Mego2vec. It is because that Mego2vec is not flexible to handle the higher-hop neighbors in matching graph, which is proven to be necessary in our experiments.
	\item  At last, our proposed method significantly outperforms the baselines in all cases. The results show that our approach is effective for addressing matching collisions in anchor link prediction by constructing matching graph, modeling both similarity of user characteristics and local structure consistency.
\end{itemize}
\subsection{Ablation Study}
In this section, we explore the effectiveness of the different numbers of hops and different numbers of sampled neighbors.  
\subsubsection{Number of hops} An important hyper-parameter in our model is the number of layers $k$, i.e. the neighbors of nodes in $k$ hops aggregated in predicting anchor links. We evaluate the performance on three datasets when only 1-hop and both 1- and 2-hop neighbors are adopted in our model. We observe that compared to the sampling with 1-hop, the MRR increases 17\%, 8\% and 1\% if we use 1 and 2 hop sampling method on three datasets when training ratio is 10\%. With more hops of neighbors, our model captures higher orders of proximity in matching graph. Compared to the Flickr-Myspace dataset, the improvement of MRR in Online-Offline is greater. It is because that in Online-Offline data, the ratio of nodes holding local structure consistency in 2-hop neighbors is more than 34\%, which is much more than the Flickr-Myspace dataset of which the ratio is about 3\%. In general, the 2-hop neighbors together with 1-hop neighbors help to avoid the matching collisions problem. In the real world data, \textit{local structure consistency} of higher hops ($k \geq 3$) seldom exists, and the number of layers of our model is set to 2. 

%
\subsubsection{Sampling Strategy} We sample the neighbors by two strategies:  random sampling and feature importance sampling. However, the result shows that these two strategies are almost the same in terms of the evaluation metrics on all  datasets. This may due to the nodes with similar attributes are not anchor nodes in practice, which we call ``matching collisions". Sampling these nodes introduces no gain for our model. For simplicity, we adopt random sampling to sample the neighbors for convolution.

\begin{figure*}[htbp]
	\centering
	\subfigure[Onlie-Offline]{
		\label{mrr_runtime:online-offline}
		\begin{minipage}[t]{0.33\linewidth}
			\centering
			\includegraphics[scale=0.32]{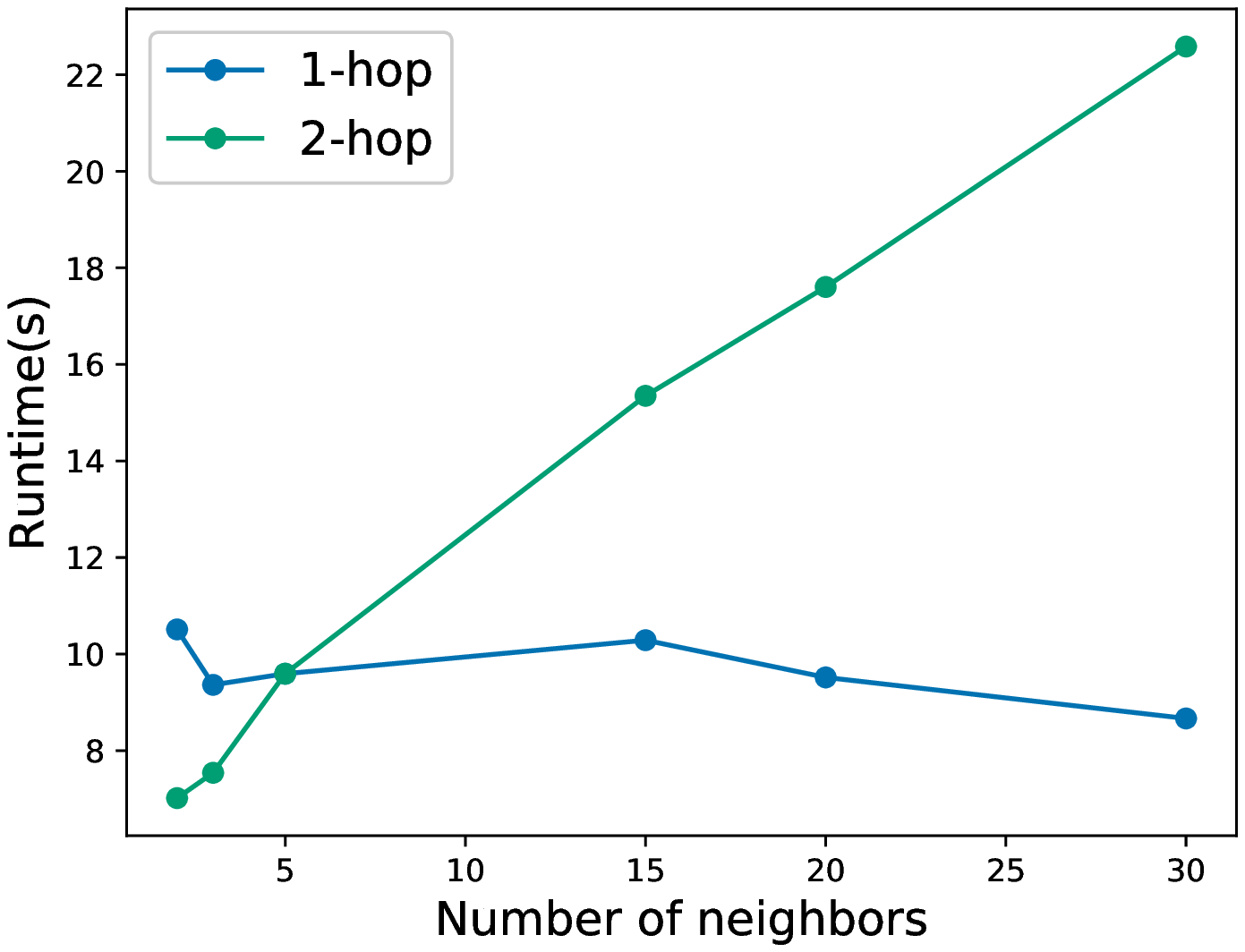}\\
			
		\end{minipage}%
	}%
	\subfigure[Flickr-Lastfm]{\label{mrr_runtime:flickr-lastfm}
		\begin{minipage}[t]{0.33\linewidth}
			\centering
			\includegraphics[scale=0.32]{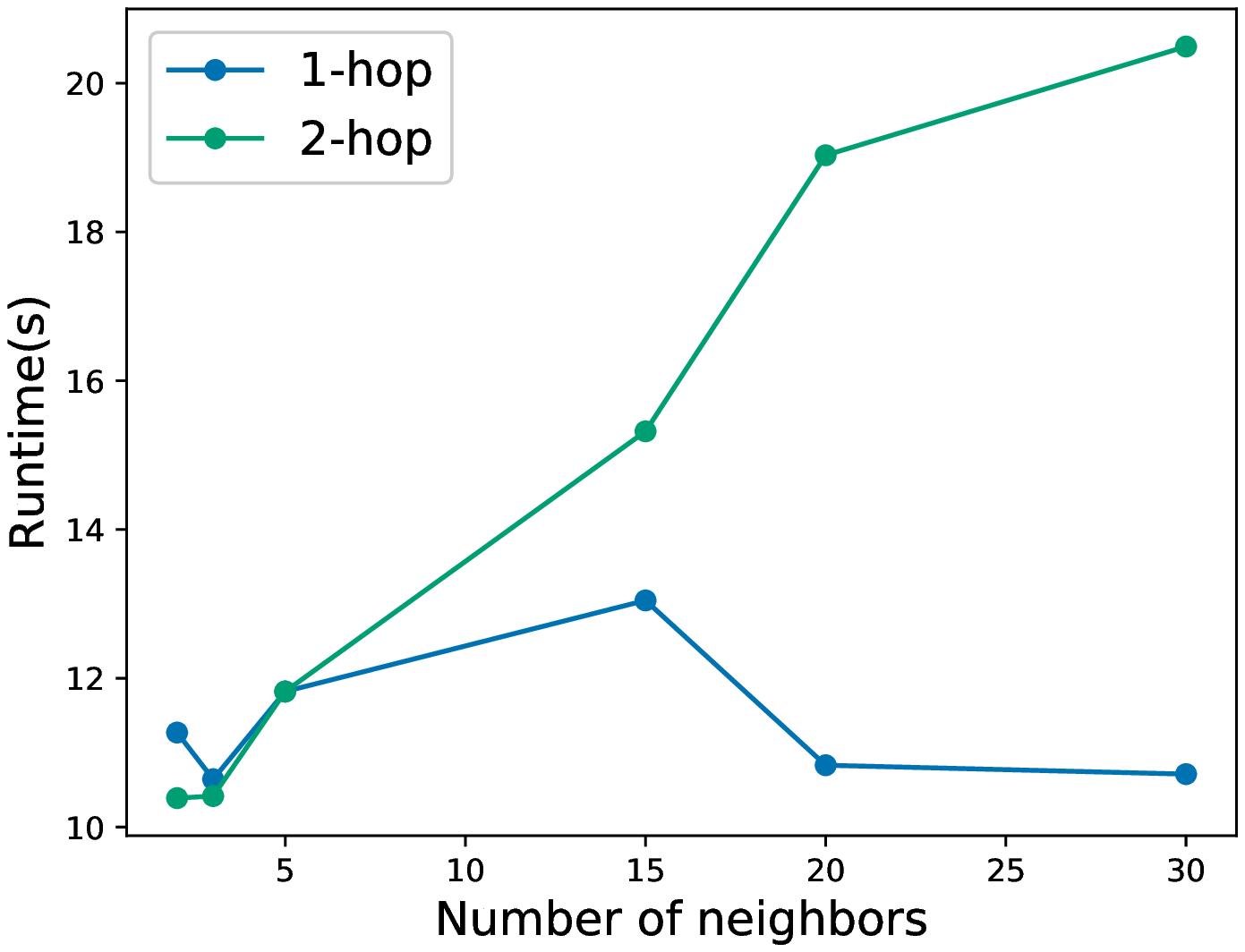}\\

		\end{minipage}%
	}%
	\subfigure[Flickr-Myspace]{\label{mrr_runtime:flickr-myspace}
		\begin{minipage}[t]{0.33\linewidth}
			\centering
	
			\includegraphics[scale=0.32]{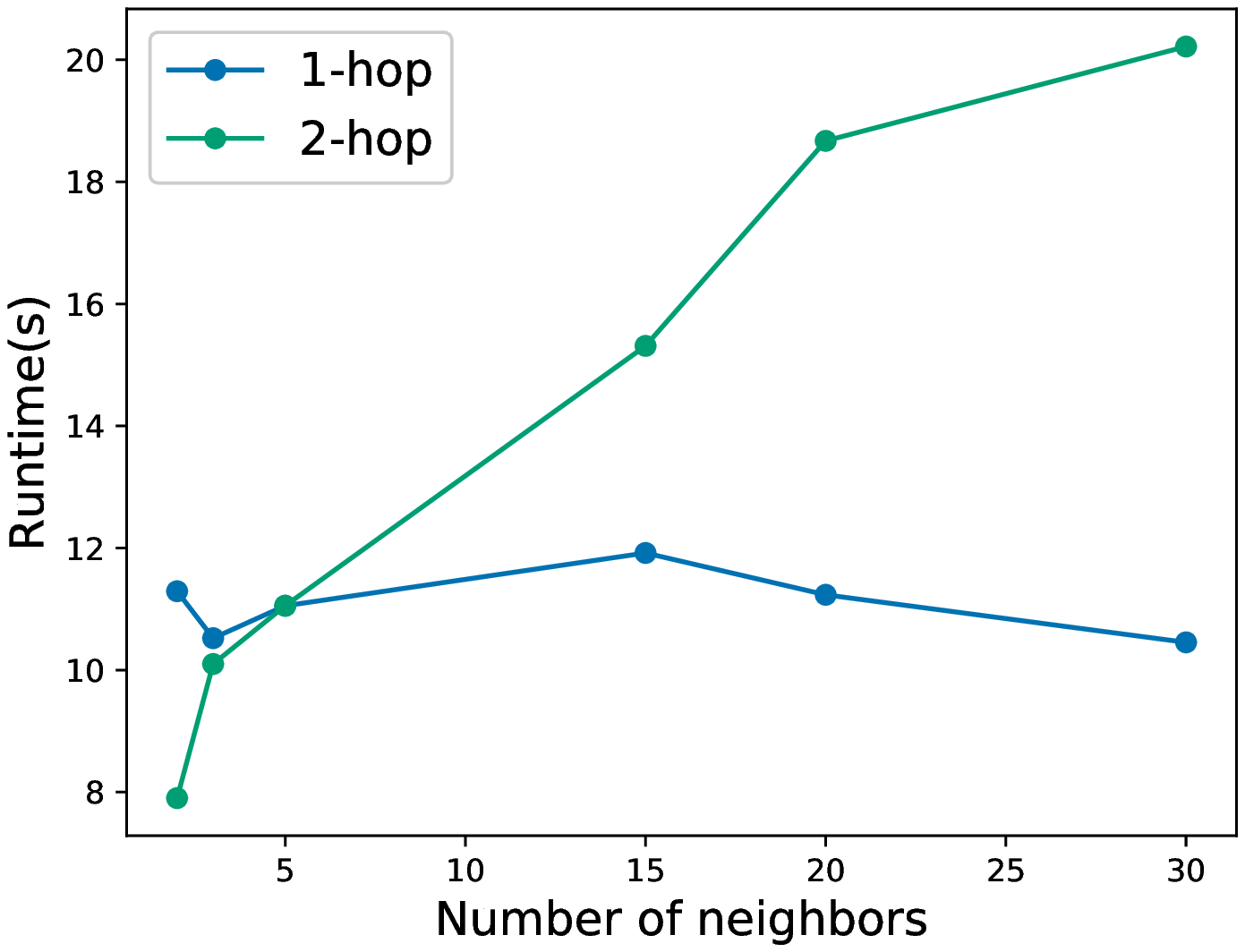}\\
			
		\end{minipage}%
	}%
	
	\centering
	\caption{Runtime of different numbers of 1 and 2 hop of samping method on different datasets.}
	\vspace{-0.2cm}
	\label{fig:mrr and runtime}
\end{figure*} 

\subsection{Analysis on Efficiency}
We study the effect of numbers of neighbors in the first and second hops. We set the number of neighbors of the first layer as 2, 3, 5, 15, 20, 30 with the fixed number of neighbors of the second layer as 5 to see how it affects the running time. Furthermore, we fix the number of neighbors in 1-hop neighbors and the number of neighbors in 2-hop ranges from 2 to 30. The result is illustrated in Figure~\ref{fig:mrr and runtime}. We observe that the running time is highly affected by the 2 hop of sampling neighbors compared to that of the 1 hop. This is because sampling 2 hop of neighbors increases the parameters quadratically. 

Besides, our model is scalable when networks become larger. Compared to the original graph convolution networks with the matching graph, our model converges within seconds under the varied scales of networks ranging from 3k to 15k. The input of the original GCNs needs to construct the matching graph, which is time and space consuming. However, our mini-batch strategy which samples nodes is efficient and the number of parameters is only relative to the complexity of aggregator function. Furthermore, to show that our proposed mini-batch way is scalable with large networks, we further explore the running time of training and testing on different sizes of constructed datasets. We first construct six random networks with node sizes from 10 to 1,000,000 by 10 times. The source and target networks are the same for simplicity. We run our model on the datasets, and both training and testing time are \textbf{log-linear} in terms of the node size of a single network.

\subsection{Embedding Visualization} 
\subsubsection{Discriminative node embeddings}

\begin{figure}[h]
	\centering
	\includegraphics[scale=0.39]{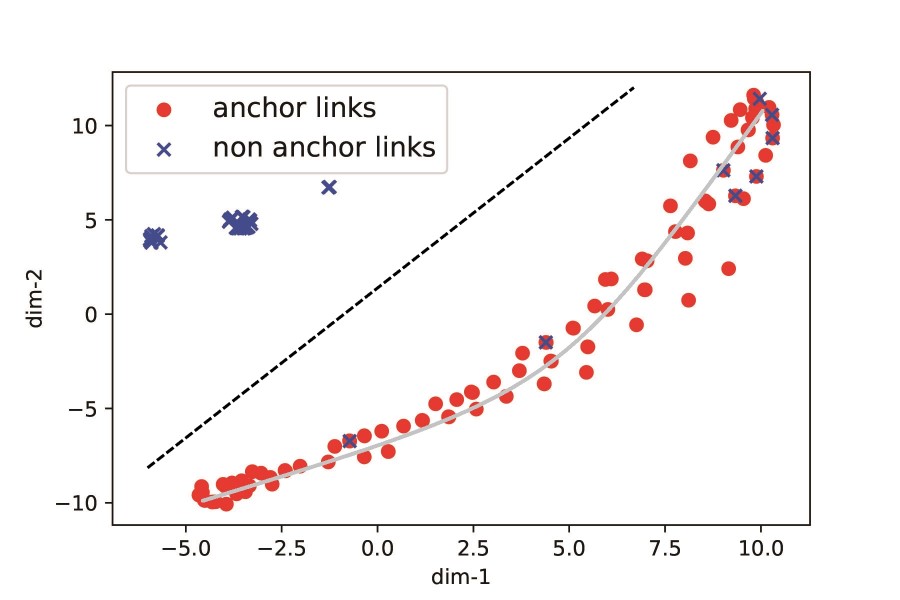}
	\caption{Effectiveness of the number of layers. Red points are anchor links, and blue crosses are non-anchor links.} \label{boundary_line}
\end{figure}

Generally, the learned node representations should directly reflect the similarity on anchor links across networks. Thus, we visualize the node embeddings on matching graph with all labeled anchors and sampled false candidates in order to check if the embeddings between anchor links and non-anchor links are distinct from each other. We implement the visualization on all observed anchor links with sampled non-anchor links in D1 by t-SNE~\cite{Rauber2016} and the results are shown in Figure~\ref{boundary_line}. We can observe a boundary between anchor links and non-anchor links, demonstrating that the learned node embeddings from our method are discriminative for finding anchor links among matching collisions.
\subsubsection{Local structure consistency}
\begin{figure}[htbp]
	\centering
	
	\subfigure[Sub-matching graph of center node.]{
		\label{ne_network}
		\begin{minipage}[t]{0.5\linewidth}
			\centering
			\includegraphics[width=1.651in]{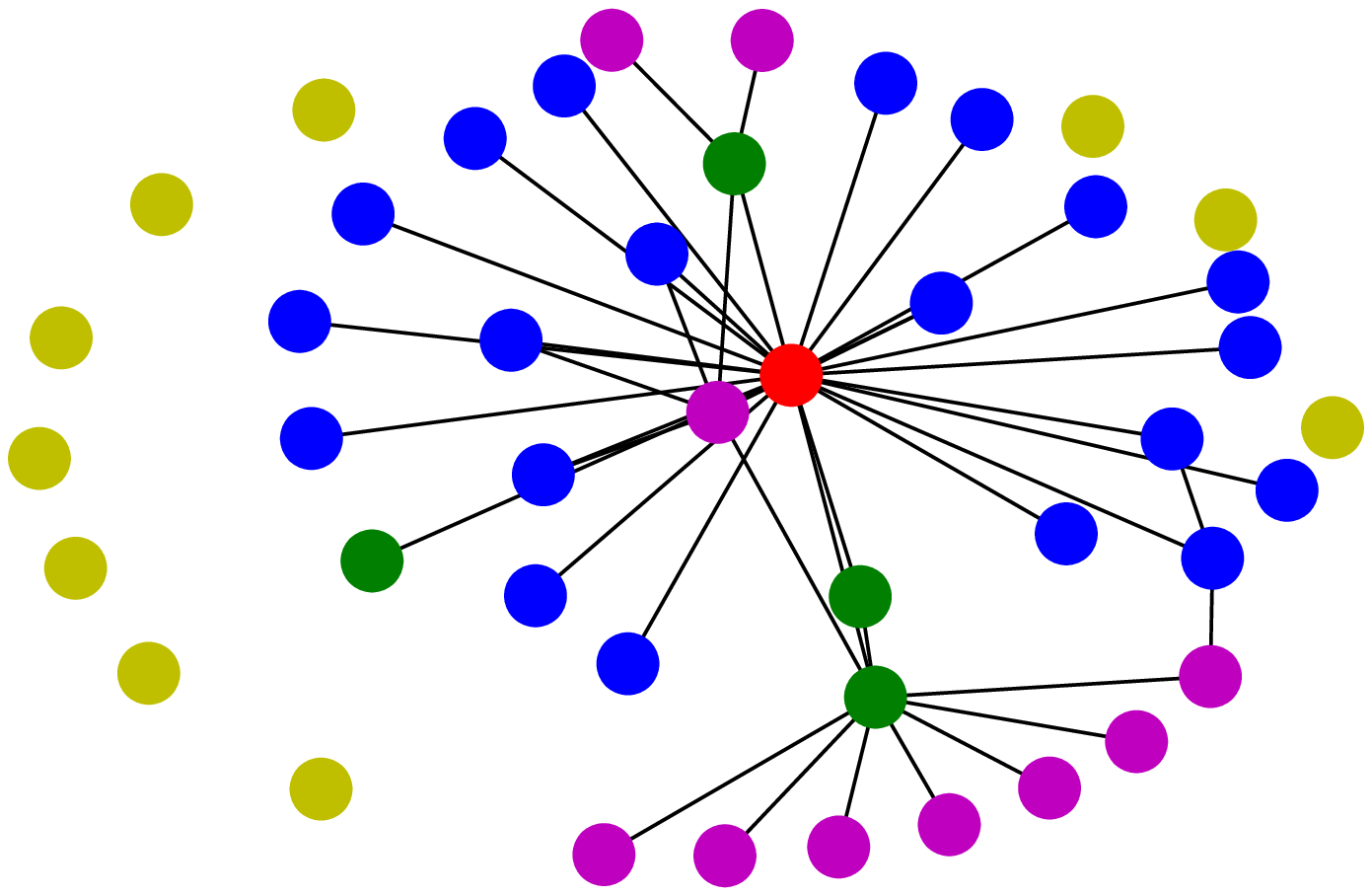}\\
			
		\end{minipage}%
	}%
	\subfigure[Embedding visualization.]{
		\begin{minipage}[t]{0.5\linewidth}
			\centering
			\includegraphics[width=1.651in]{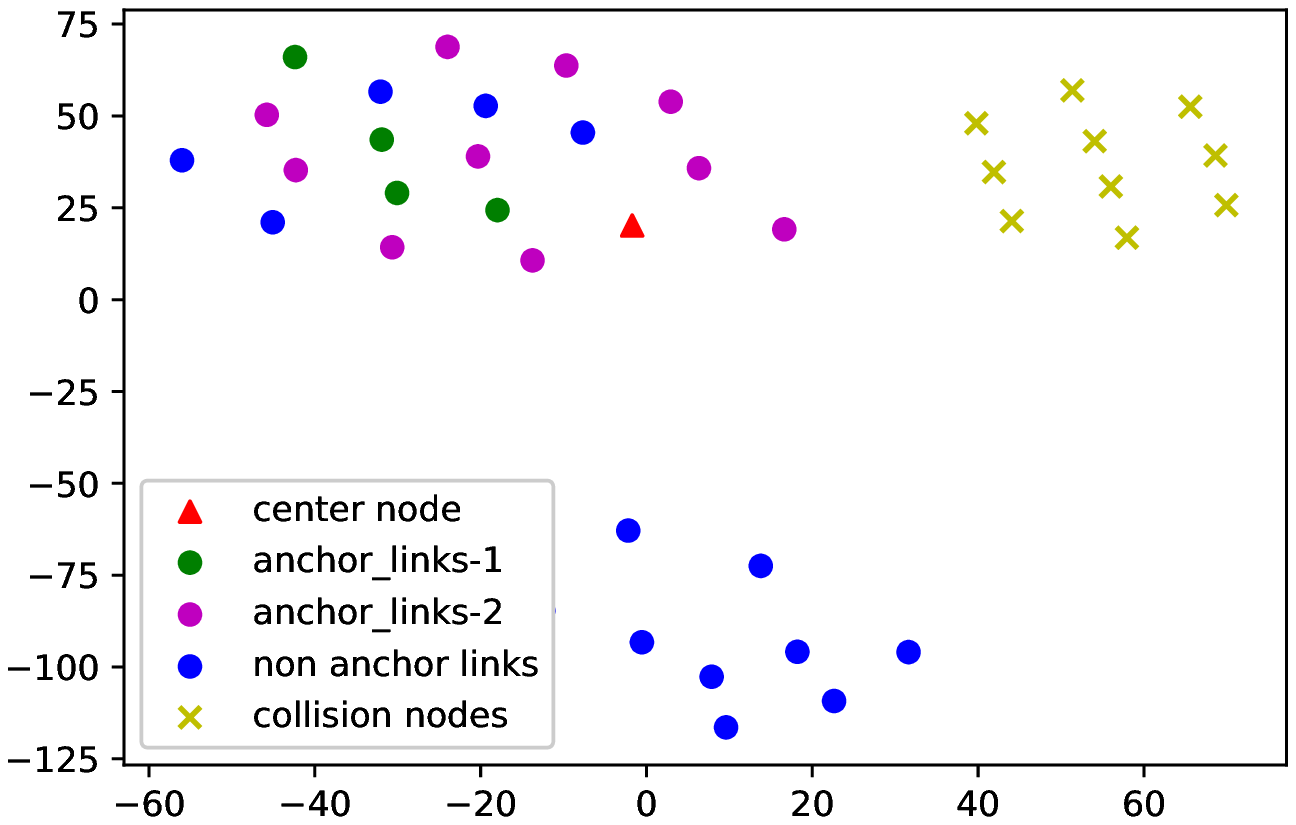}\\
		\end{minipage}%
	}%
	
	\centering
	\caption{Matched graph of randomly selected nodes and the embedding visualization.}
	\vspace{-0.2cm}
	\label{fig:network and embedding}
\end{figure}

According to the visualization on node embeddings in Figure~\ref{boundary_line}, we find that there exists a chain connecting anchor links on matching graph. We wonder if the node embeddings are relative to each other according to the matching edges on the matching graph. Therefore, we implement visualization on a subgraph from full matching graph by t-SNE~\cite{Rauber2016}.  The structure of the subgraph is shown in Figure~\ref{fig:network and embedding}(a), where the nodes colored by red, green, purple and yellow are anchor links. In the subgraph, the red node is the center node, the green and blue ones are 1-hop neighbors of the center node and the purple ones are 2-hop neighbors. Note that the yellow nodes are the nodes far away from the center node on matching graph. The visualization of node embeddings on the subgraph is depicted in Figure~\ref{fig:network and embedding}(b). We observe that there are three clusters in the visualization space.  The first cluster (upper left) mainly consists of the center node and its 1- and 2-hop neighbors labeled by anchor links. The second cluster (lower left) consisting of 1-hop neighbors labeled by non-anchor links, is separated from the first cluster, which also indicates the discriminative node embeddings learned by our method. Moreover, the third cluster (upper right) consisting of anchor links far away from the center nodes, is also separated from the first cluster. The results imply that the learned node embeddings contain local structure consistency across networks. In this way, the inference of anchor links should be followed by both similarity
of user characteristics and the closeness to labeled anchor links, which exactly verify our motivation to solve match collisions. 
\section{Conclusion}\label{Conclusion}
In this paper, we proposed a novel method using GCN to predict anchor links on matching graph solving matching collisions. We introduced a scalable algorithm to train GCN networks. Then we evaluated our method on three real-world datasets, and the result demonstrated that our method outperforms the comparison methods. In the future, we aim to study how to predict anchor links on multiple networks.
\section*{Acknowledgment}
This work was funded by the National Natural Science Foundation of China under grant numbers 61802371, 91746301, the National Key Research and Development Program of China under grant numbers 2018YFC0825200, 2016QY03D0504 and the National Social Science Fund of China under grant number 19ZDA329. Yongqing Wang is also funded by CCF-Tencent Open Research Fund under grant number RAGR20190117.

\bibliographystyle{IEEEtran}
\bibliography{bibi}

\end{document}